\documentclass[10pt]{article}
\usepackage[utf8]{inputenc}
\DeclareUnicodeCharacter{2212}{−-}
\usepackage{graphicx}
\usepackage{hyperref}
\usepackage{url}
\usepackage{amsmath}
\usepackage{siunitx}
\usepackage{longtable,tabularx}
\usepackage{nccmath}
\usepackage{amsmath}
\usepackage[superscript, nomove]{cite}
\usepackage[margin=1in]{geometry}
\usepackage[toc,page]{appendix}
\usepackage{setspace}
\usepackage{amssymb}
\usepackage{tabu}
\usepackage{listings}
\usepackage{afterpage}
\usepackage{titlesec}
\usepackage[labelfont=bf]{caption}
\usepackage{fancyhdr}
\usepackage{pgf}
\usepackage{float}
\usepackage{subcaption}
\usepackage{cleveref}
\usepackage{textcomp}
\usepackage{pict2e}
\usepackage{wasysym}
\usepackage{multicol}
\usepackage{lipsum}
\usepackage{indentfirst}
\usepackage{amsmath}
\usepackage{amssymb}
\usepackage{mathtools}
\usepackage{amsthm}
\usepackage{xspace}
\usepackage{subcaption}
\usepackage{caption}
\usepackage{microtype}
\usepackage{graphicx}
\usepackage{booktabs}
\usepackage{tcolorbox}
\usepackage{multirow}
\usepackage{xcolor,colortbl}

\titleformat{\section}
{\normalfont\normalsize\bfseries}{\thesection}{0.5em}{}
\titleformat{\subsection}
{\normalfont\normalsize\bfseries\em}{\thesubsection}{0.5em}{}
\titleformat{\subsubsection}
{\normalfont\normalsize\bfseries\em}{\thesubsubsection}{0.5em}{}

\usepackage{amsmath,amsfonts,bm}
\usepackage{mathtools}

\def\eqref#1{equation~\ref{#1}}

\def\1{\bm{1}}

\DeclareMathAlphabet{\mathsfit}{\encodingdefault}{\sfdefault}{m}{sl}
\SetMathAlphabet{\mathsfit}{bold}{\encodingdefault}{\sfdefault}{bx}{n}

\newcommand{\E}{\mathbb{E}}

\newcommand{\Eb}[2]{\E_{#1}\!\left[#2\right]}

\newcommand{\bx}{\mathbf{x}}

\newcommand{\bepsilon}{{\boldsymbol{\epsilon}}}

\definecolor{tabfirst}{rgb}{1, 0.7, 0.7} 
\definecolor{tabsecond}{rgb}{1, 0.85, 0.7} 
\definecolor{tabthird}{rgb}{1, 1, 0.7} 

\begin{document}

\begin{flushright}
\Large 

\textbf{[SSC24-X-04]}
\end{flushright}
\begin{centering}      
\large 

\textbf{Beyond the Visible: Jointly Attending to Spectral and Spatial Dimensions with HSI-Diffusion for the FINCH Spacecraft}\\
\vspace{0.5cm}
\normalsize 

{\textbf{Ian Vyse, Rishit Dagli, Dav Vrat Chadha, John P. Ma, Hector Chen, Isha Ruparelia, Prithvi Seran, Matthew Xie, Eesa Aamer, Aidan Armstrong, Naveen Black, Ben Borstein, Kevin Caldwell, Orrin Dahanaggamaarachchi, Joe Dai, Abeer Fatima, Stephanie Lu, Maxime Michet, Anoushka Paul, Carrie Ann Po, Shivesh Prakash, Noa Prosser, Riddhiman Roy, Mirai Shinjo, Iliya Shofman, Coby Silayan, Reid Sox-Harris, Shuhan Zheng, Khang Nguyen}}\\
{University of Toronto Aerospace Team - Space Systems\footnote{University of Toronto Aerospace Team is a student group affiliated with The University of Toronto, Toronto, Canada.}}\\
{55 St. George Street, Myhal Centre, Room 618, Toronto, ON}\\
{\href{mailto:ss-outreach@utat.ca}{\texttt{ss-outreach@utat.ca}}}

\end{centering}

\begin{figure*}[!h]
    \centering
    \begin{tabular}{cccc}
        \multicolumn{2}{c}{\textbf{Band $100$}} &  \multicolumn{2}{c}{\textbf{Band $220$}}\\
        \textbf{Striped Image} & \textbf{Ours (HSI-Diffusion)} & \textbf{Striped Image} & \textbf{Ours (HSI-Diffusion)}\\
        \midrule
        \includegraphics[width=0.23\textwidth]{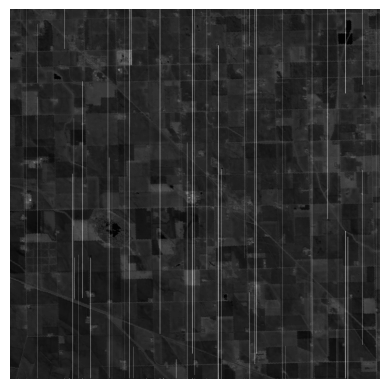} & \includegraphics[width=0.23\textwidth]{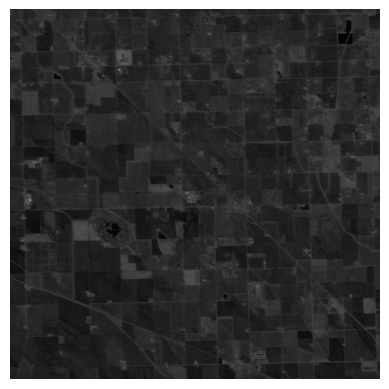} & \includegraphics[width=0.23\textwidth]{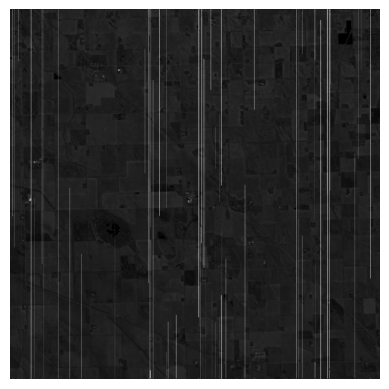} & \includegraphics[width=0.23\textwidth]{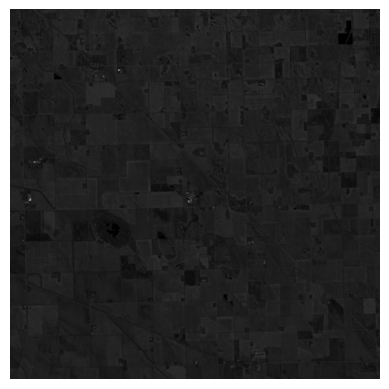} \\
        \includegraphics[width=0.23\textwidth]{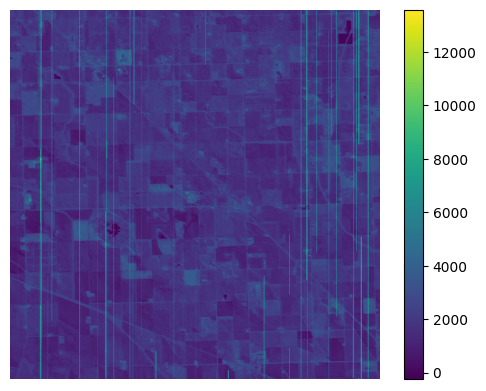} & \includegraphics[width=0.23\textwidth]{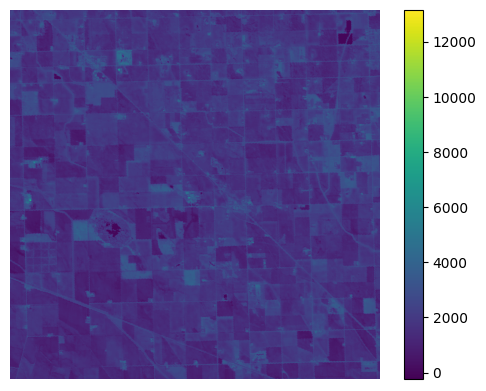} & \includegraphics[width=0.23\textwidth]{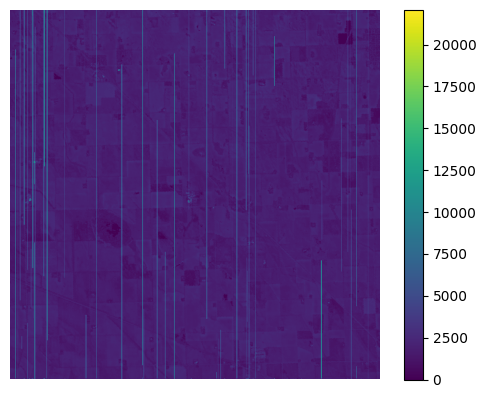} & \includegraphics[width=0.23\textwidth]{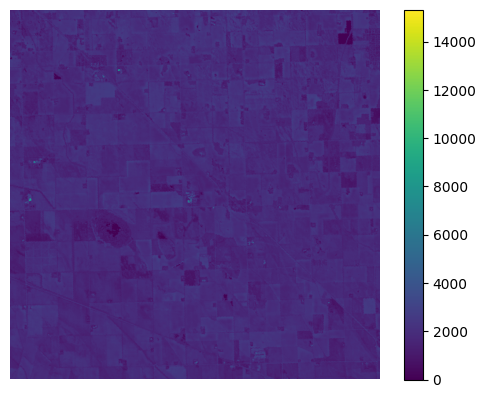}
    \end{tabular}
    \caption{\emph{Hyperpsectral Destriping.} We present Hyperspectral Diffusion, a technique that can denoise or destripe satellite hyperspectral data cubes. We demonstrate results from real collected data from the EnMAP hyperspectral satellite mission~\cite{STORCH2023113632,rs70708830} which is analogous to images we expect to be captured from our FINCH satellite. Best viewed with color and zoom.}
    \label{fig:teaser}
\end{figure*}
\begin{centering}
    \vspace{0.5cm}
    \centerline{\textbf{ABSTRACT}}
    \vspace{0.3cm}
\end{centering}

\definecolor{pk}{rgb}{0.988, 0.286, 0.639}

Satellite remote sensing missions have gained popularity over the past fifteen years due to their ability to cover large swaths of land at regular intervals, making them ideal for monitoring environmental trends. The FINCH mission, a 3U+ CubeSat equipped with a hyperspectral camera, aims to monitor crop residue cover in agricultural fields. Although hyperspectral imaging captures both spectral and spatial information, it is prone to various types of noise, including random noise, stripe noise, and dead pixels. Effective denoising of these images is crucial for downstream scientific tasks. Traditional methods, including hand-crafted techniques encoding strong priors, learned 2D image denoising methods applied across different hyperspectral bands, or diffusion generative models applied independently on bands, often struggle with varying noise strengths across spectral bands, leading to significant spectral distortion. This paper presents a novel approach to hyperspectral image denoising using latent diffusion models that integrate spatial and spectral information. We particularly do so by building a 3D diffusion model and presenting a 3-stage training approach on real and synthetically crafted datasets. The proposed method preserves image structure while reducing noise. Evaluations on both popular hyperspectral denoising datasets and synthetically crafted datasets for the FINCH mission demonstrate the effectiveness of this approach.\\ \\
Our code can be found at \href{https://github.com/utat-ss/FINCH-destriping}{\texttt{\textcolor{pk}{github.com/utat-ss/FINCH-destriping}}}

\begin{multicols*}{2}
\section{INTRODUCTION}

The FINCH spacecraft~\cite{miles2022finch, murdoch2023finch} (Field Imaging Nanosatellite for Crop residue Hyperspectral mapping) represents a significant advancement in the field of agricultural monitoring via remote sensing technologies. Currently under development by the University of Toronto Aerospace Team's Space Systems division, this 3U+ CubeSat is designed to precisely estimate the percentage of crop residue cover across Canadian agricultural fields employing advanced spectral unmixing techniques facilitated by its onboard hyperspectral remote sensing payload. The spacecraft integrates a pushbroom sensor~\cite{615446,7025137}, necessitating a robust data processing pipeline to transform raw spectral data into scientifically usable information.

Hyperspectral imaging has been pivotal in remote sensing, providing detailed spectral information across a plethora of contiguous bands, and enabling precise analysis of the Earth’s surface or, in the case of the FINCH mission, crop residue mapping. This technology continues to become more accessible as an ``image-based acquisition tool for physically meaningful measurements"~\cite{y15}. Hyperspectral denoising, critical to ensuring that hyperspectral data is scientifically useful, has undergone significant evolution over the years, moving from traditional image-processing techniques~\cite{6165657,He_2019_CVPR} to advanced deep-learning approaches~\cite{8913713,9318503,6819824,8693549,8454887}. However, the deployment of hyperspectral sensors, such as the pushbroom sensor~\cite{615446,7025137} aboard FINCH, introduces inherent challenges, notably the presence of striping and random noise~\cite{Gomez-Chova:08,rogass2014reduction}. Striping noise, which results in linear distortions along the sensor's scanning direction, is akin to artifacts observed in other hyperspectral satellites like EO-1 Hyperion~\cite{EO1Hyperion}. This type of noise, if unaddressed, can severely compromise the spatial and spectral integrity of the data~\cite{He_2019_CVPR,Zhang_2021_ICCV,8913713,6165657}, rendering it less reliable for scientific analysis. The removal of striping noise, caused by the operation of the push-broom sensor, is particularly relevant for FINCH, where stripes can compromise both the spatial and spectral integrity of the data collected.

Addressing the striping noise challenge is critical for performing any science on images from the FINCH mission. We present a 3D hyperspectral diffusion model designed to effectively mitigate striping noise. We particularly do so by building a 3D diffusion model and perform a 3-stage training. We train our approach on real and synthetic data. We also compare our approach with existing denoising models, where we find that our approach not only demonstrates superior performance in destriping on satellite images but also shows potential for broader inpainting applications within remote sensing.

The implementation of the hyperspectral diffusion destriping model within the FINCH spacecraft's operational schema marks a pivotal development in remote sensing technology. By effectively addressing the issue of striping noise in hyperspectral data, this model significantly improves the data quality, thereby enhancing the reliability of remote sensing analyses. The application of deep learning techniques in the noise reduction process not only showcases the model's technical prowess but also expands the potential for these methodologies in refining data accuracy across various remote sensing applications. Moreover, the scalability of this solution presents opportunities for integration into additional satellite systems, potentially broadening its impact on the field. 

\paragraph{Contributions.} Our contributions can be summarized as follows,
\begin{itemize}
    \item We present a hyperspectral diffusion destriping model to remove the various types of striping noise that can occur within FINCH data. Our approach can significantly enhances the quality of data by mitigating striping noise, which is critical for maintaining both the spatial and spectral integrity of hyperspectral remote sensing data.
    \item We perform comprehensive comparisons with existing denoising models, demonstrating that the proposed model performs favorably in removing various types of synthetic and real noise from hyperspectral data.
\end{itemize}

\section{RELATED WORKS}

Noise is a common issue in remote sensing imagery, often appearing in the form of stripes due to factors such as poor sensor calibration~\cite{4215084}. These striping artifacts degrade data quality, cause poor interpretability, and may also cause failures in image processing and analysis systems~\cite{9234507,7433945,7542167,rs13040827,9376777,9826537,PANDECHHETRI2011620,8734833,4683341,9552462}. Therefore, it is highly valuable to develop algorithms for identifying and correcting or at least alleviating striping noise~\cite{4683341}. Dead pixel correction is an additional challenge in hyperspectral images (HSI) as dead pixels can significantly affect the data quality of hyperspectral data. Methods that detect and correct these issues ensure the continuity of the spectral information across the image, thereby enhancing the overall quality of the data. 

Many traditional approaches to denoising hyperspectral images are statistical-based methods that have been used for denoising typical triband images. Some of these approaches includes using wavelet transformations~\cite{PANDECHHETRI2011620}, low-rank decomposition~\cite{8533598} , total variation~\cite{7044559}, and sparse representation~\cite{10.1007/978-3-319-46478-7_2}. Although these methods have shown some effectiveness, they are limited by their tendency to generalize to very specific and varying types of stripes that may occur in the model. This generalization can make it difficult to achieve proper restoration across a multitude of cases, leading to the creation of large, complex statistical models that lack interpretability and are difficult to reuse. 

Machine learning approaches, particularly those involving convolutional neural networks (CNNs), such as HSI-DeNet~\cite{8435923}, have recently been explored for restoring hyperspectral images. Within this context, various approaches have been explored to achieve the denoising task, such as decomposing the task into multiple types of stripes~\cite{PAN2023109832}, as well as employing wavelet transforms~\cite{10189870}. Another significant approach is the Multi-scale Adaptive Feature Network (MAFNet)~\cite{rs12050872}. This method is designed to handle varying noise levels and structures by adaptively learning from the data to identify a complex mapping between noisy and clean hyperspectral imaging~\cite{pgdd23}. GLCSA-Net~\cite{GLCSA}, which employs a generative adversarial network (GAN) as the backbone, has been developed for describing and denoising based on deep prior models~\cite{chen2024}.  It is important to note that hyperspectral images contain low-level information, which has been demonstrated to be present in both two-dimensional and three-dimensional convolutional networks. This information can serve as prior information in various other models~\cite{DBLP:journals/corr/abs-1902-00301}.

As transformer-based architectures~\cite{vaswani2023attention} have been explored for various applications, it is intuitive that many have also considered using them for hyperspectral image (HSI) denoising. The Mixed Attention Network (MAN)~\cite{lai2023mixed} proposes methods to dynamically control the flow of information through a transformer-based model. One problem with hyperspectral images is information imbalance among spectral-spatial bands. By applying attention to assign importance to skip connections and spectral bands, the model learns to become extremely adept at interpreting complicated hyperspectral feature spaces. The Hyperspectral Denoising Transformer (HSDT)~\cite{lai2023hsdt} builds upon the idea of attentive information flow by modifying the attention blocks to look directly at the full 3D spectral-spatial space. This allows the model to generalize much better than previous models. Several advanced deep learning models employed for tasks such as image super-resolution and reconstruction share various degrees of similarity with hyperspectral denoising. The Spectral Enhanced Rectangle Transformer (SERT)~\cite{nejad2023sert} uses transformers to handle complex noise patterns in HSIs.
  
Recently, diffusion has been used in various tasks, such as denoising~\cite{ho2020denoising,NEURIPS2022_95504595,Zhu_2023_CVPR,pmlr-v202-kulikov23a,dagli2023diffuseraw}, generating images from text~\cite{NEURIPS2019_1d72310e,Rombach_2022_CVPR,zhang2023adding,pmlr-v139-ramesh21a}, and inpainting~\cite{lugmayr2022repaint}. These methods have also spread to hyperspectral image denoising. Models such as SpectralDiff~\cite{Chen_2023} have used diffusion as an approach to denoise hyperspectral images. However, it is essential to recognize that it uses 2D convolutions instead of 3D. With the three-dimensional nature of hyperspectral images, where multiple bands can provide cross-information of the noise in the image, it is possible to fully exploit the 3D data in service of the denoising task. This could involve developing models that can simultaneously leverage information across all bands, improving the overall destriping and denoising performance by utilizing the inherent spectral correlations present in hyperspectral images. In this work, we propose a 3D diffusion model to tackle this problem. 

\begin{figure*}[t]
    \centering
    \includegraphics{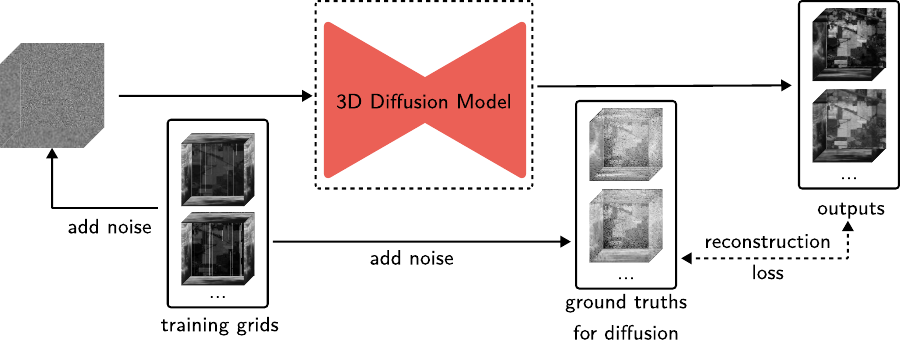}
    \caption{Our method trains a 3D diffusion model across 3 stages to perform destriping on hyperspectral satellite images like the ones that would be captured by FINCH.}
    \label{fig:methods}
\end{figure*}

\section{PRELIMINARY: DIFFUSION MODELS}

Diffusion models~\cite{sohl2015deep, ho2020denoising, song2019generative} are a class of generative models that iteratively transform noise into structured data through a sequence of denoising steps. Unlike traditional generative models that directly model the data distribution, diffusion models utilize a forward and backward diffusion process to generate data. Specifically, the forward diffusion process gradually adds noise to the input data over a series of time steps, creating increasingly noisy versions of the data. Given an initial data point $\bx_0$, the forward process produces $\bx_t$ by adding Gaussian noise in t discrete steps:

\begin{equation}
q(\bx_t | \bx_{t-1}) = \mathcal{N}\left(\bx_t; \sqrt{1 - \beta_t} \bx_{t-1}, \beta_t I\right), 
\end{equation}

where $\beta_t$ is a variance schedule that controls the amount of noise added at each step. The backward diffusion process, also known as the denoising process, reverses this noising process. It aims to reconstruct the original data from the noisy versions. This is achieved by training a neural network $\epsilon_\theta$ to predict the noise added at each step. The model is trained to minimize the difference between the predicted noise and the actual noise added in the forward process, often using a mean squared error loss:
\begin{equation}
\Eb{\bx_0, t, \bepsilon \sim \mathcal{N}(0, I)}{\| \bepsilon - \epsilon_\theta(\bx_t, t) \|^2_2}, 
\end{equation}

where $\bx_t$ is the noisy image at time step t, and $\bepsilon$ is the added Gaussian noise. During inference, the model starts with a random noise $\bx_T$ and iteratively denoises it using the trained model $\epsilon_\theta$, producing a data sample $\bx_0$ that resembles the training data.

\section{METHODS}

We next describe our approach to train our diffusion model (\Cref{sec:training}) and create the synthetic training data (\Cref{sec:synthetic}).

\subsection{Training the Diffusion Model}
\label{sec:training}

Following recent progress in generative models~\cite{sohl2015deep, song2019generative, ho2020denoising, Rombach_2022_CVPR}, we build a 3D diffusion model for hyperspectral image destriping that operates on $32 \times 32 \times 32$ occupancy grids of hyperspectral images. To do so, we adopt a three-stage training approach, which we describe. Initially, we adapt a U-Net architecture, previously employed for the Pavia University~\cite{granahyperspectral} dataset's 2D image segmentation, to support 3D hyperspectral input. This adaptation includes an expansion of the model's convolutional layers to process an additional spectral dimension, thus accommodating the intricate characteristics of hyperspectral images. Subsequently, the model undergoes a denoising training phase using the ICVL-HSI~\cite{10.1007/978-3-319-46478-7_2} dataset, which reflects diverse real-world noise conditions, excluding satellite hyperspectral imagery. The final stage involves fine-tuning our model on a synthetic dataset derived from multiple hyperspectral scenes, ensuring robust performance across various spectral scenarios. We illustrate this in~\Cref{fig:methods}.

\footnotetext{We were unable to reproduce results from most of the methods and we take metrics directly from the papers.}
\begin{table*}[ht]
    \centering
    \caption{\emph{Results on Real Data.} We report the PSNR, SSIM, SAM, and LPIPS metrics for different models$^{1}$. The \colorbox{tabfirst}{best}, \colorbox{tabsecond}{second best}, and \colorbox{tabthird}{third best} results for each metric are color coded.}
    \label{tab:real_results}
    \begin{tabular}{p{2cm}ccccc}
        \toprule
        \textbf{Modification} & \textbf{Model} & \textbf{PSNR} $\uparrow$ & \textbf{SSIM} $\uparrow$ & \textbf{SAM} $\downarrow$ & \textbf{LPIPS} $\downarrow$ \\
        \midrule
        \multirow{7}{3em}{-} & GRN-HSI\cite{9397278} & 40.58 & 0.9840 & 0.0730 & - \\
        & Deep HS Prior 3D\cite{DBLP:journals/corr/abs-1902-00301} & 23.24 & 0.8520 & 9.9100 & - \\
        & MAFNet\cite{pgdd23} & 42.15 & 0.9630 & \cellcolor{tabfirst}0.0290 & - \\
        & SERT\cite{li2023spectral} & \cellcolor{tabsecond} 43.68 & \cellcolor{tabfirst}0.9969 & 1.9700 & - \\
        & HSDT\cite{lai2023hybrid} & \cellcolor{tabfirst}44.69 & \cellcolor{tabsecond}0.9940 & \cellcolor{tabsecond}0.0380 & - \\
        & MAN\cite{lai2023mixed} & \cellcolor{tabthird}43.44 & \cellcolor{tabthird}0.9890 & 0.0440 & - \\
        \cmidrule{2-6}
        & \textbf{Ours (HSI-Diffusion)} & 39.2274  & 0.8817 & \cellcolor{tabthird}0.0423 & 0.2214 \\
        \midrule
        \multirow{16}{7em}{Non-i.i.d Gaussian Noise w/ $\sigma \in [0, 15]$}
        & BM4D~\cite{bm4d}          &                      44.39 &                      0.9683 &                      0.0692 & - \\
        & MTSNMF~\cite{MTSNMF}      &                      45.39 &                      0.9592 &                      0.0845 & - \\
        & LLRT~\cite{LLRT}          &                      45.74 &                      0.9657 &                      0.0832 & - \\
        & NGMeet~\cite{NGMeet}      &                      39.63 &                      0.8612 &                      0.2144 & - \\
        & LRMR~\cite{LRMR}          &                      41.50 &                      0.9356 &                      0.1289 & - \\
        & FastHyDe~\cite{FastHyDe}  &                      48.08 &                      0.9917 &                      0.0404 & - \\
        & LRTF L$_0$~\cite{LRTF}    &                      43.41 &                      0.9315 &                      0.0570 & - \\
        & E-3DTV~\cite{E3DTV}       &                      46.05 &                      0.9811 &                      0.0560 & - \\
        & T3SC~\cite{T3SC}          &                      49.68 &                      0.9912 &                      0.0486 & - \\
        & MAC-Net~\cite{macnet}     &                      48.21 &                      0.9915 &                      0.0387 & - \\
        & NSSNN~\cite{NSSNN}        &                      49.83 &                      0.9934 &                      0.0302 & - \\
        & TRQ3D~\cite{TRQ3D}        &                      46.43 &                      0.9878 &                      0.0437 & - \\
        & SST~\cite{SST}            &  \cellcolor{tabthird}50.87 &  \cellcolor{tabthird}0.9938 &  \cellcolor{tabthird}0.0298 & - \\
        & SSRT-UNET~\cite{SSRTUNET} &  \cellcolor{tabfirst}52.12 &  \cellcolor{tabfirst}0.9950 &  \cellcolor{tabfirst}0.0225 & - \\
        & SSUMamba~\cite{ssumamba}  & \cellcolor{tabsecond}51.34 & \cellcolor{tabsecond}0.9946 & \cellcolor{tabsecond}0.0256 & - \\
        \cmidrule{2-6}
        & \textbf{Ours (HSI-Diffusion)} & 38.3892 & 0.9864 & 0.0474 & 0.2274\\

        \bottomrule
    \end{tabular}
\end{table*}


\paragraph{Stage I.} We start by first reconfiguring a U-Net trained on the Pavia University dataset~\cite{granahyperspectral} for image segmentation. We expand the underlying convolutional layers by adding a channel and modifying the weights denoted as $\mathbf{W}^{2D} \in \mathbb{R}^{m \times n \times c_{in} \times c_{out}}$, into weights for the new network with an additional channel denoted as, $\mathbf{W}^{3D} \in \mathbb{R}^{m \times n \times l \times c_{in} \times c_{out}}$. We do so by replicating 2D filters to expand into the spectral domain by averaging these filters across the additional spectral dimension,
\begin{equation}
\begin{split}
    \mathbf{W}^{3D}{[:,:,k,:,:]} &= \frac{1}{l} \sum_{i=1}^{l} \mathbf{W}^{2D}_{:,:,:,:}, \\
    & \qquad \forall k \in {1, ..., l}, 
\end{split}
\end{equation}

where $l$ represents the number of spectral layers. This transformation allows us to incorporate the essential spectral information for processing hyperspectral images.

\paragraph{Stage II.} We now train the model from Stage I for denoising on the ICVL-HSI dataset~\cite{10.1007/978-3-319-46478-7_2}. The ICVL-HSI dataset~\cite{10.1007/978-3-319-46478-7_2} provides a variety of real-world scenarios reflective of common noise patterns in hyperspectral imaging. However, this dataset does not include any satellite hyperspectral images. We introduce this step to perform large-scale pre-training since our synthetic dataset is very small in size and is sampled from only $7$ hyperspectral datacubes. During this stage, we train the network as a 3D denoising diffusion model with the objective,
\begin{equation}
\mathcal{L}_{\text{denoise}}(\theta) = \mathbb{E}_{\mathbf{X}, \mathbf{N}} \left[ \left| \epsilon_{\theta}(\mathbf{X}_{\text{noisy}}, t) - \mathbf{N} \right|_2^2 \right],
\end{equation}
where $\mathbf{X}_{\text{noisy}} = \mathbf{X}_{\text{clean}} + \mathbf{N}$ and $\mathbf{N} \sim \mathcal{N}(0, \sigma^2 \mathbf{I})$.

\paragraph{Stage III.} We now fine-tune the model from Stage II using a synthetic dataset derived from the Hyperspectral Remote Sensing Scenes database~\cite{granahyperspectral} comprised of the Indian Pines~\cite{PURR1947}, Salinas~\cite{granahyperspectral}, Pavia Centre~\cite{granahyperspectral}, Pavia university~\cite{granahyperspectral}, Cuprite~\cite{granahyperspectral}, KSC~\cite{granahyperspectral}, and Botswana~\cite{granahyperspectral} datasets. We describe the creation of this synthetic dataset in~\Cref{sec:synthetic}. We fine-tune the 3D denoising diffusion model from stage II using the following objective,
\begin{equation}
\mathcal{L}_{\text{fine-tune}}(\theta) = \mathbb{E}_{\mathbf{X}, \mathbf{N}} \left[ \left| \epsilon_{\theta}(\mathbf{X}_{\text{noisy}}, t) - \mathbf{N} \right|_2^2 \right]. 
\end{equation}

\begin{table*}[t]
    \centering
    \caption{\emph{Results on Synthetic Data.} We report the PSNR, SSIM, SAM, and LPIPS metrics for various methods tested on synthetic datasets for the FINCH mission. The \colorbox{tabfirst}{best}, \colorbox{tabsecond}{second best}, and \colorbox{tabthird}{third best} results for each metric are color coded.}
    \label{tab:syn_results}
    \begin{tabular}{p{4cm}cccccccc}
        \toprule
        \textbf{Method} & \multicolumn{4}{c}{\textbf{Pavia}} & \multicolumn{4}{c}{\textbf{WDC}} \\
        \cmidrule(lr){2-5} \cmidrule(lr){6-9}
                        & \textbf{PSNR} & \textbf{SSIM} & \textbf{SAM} & \textbf{LPIPS} & \textbf{PSNR} & \textbf{SSIM} & \textbf{SAM} & \textbf{LPIPS} \\
        \midrule
        \textbf{BM4D\cite{bm4d}}    & 25.0946       & 0.6944        & 0.2325       & -              & 24.5261       & 0.5675        & 0.2198       & - \\
        \textbf{LRMR\cite{LRMR}}    & 31.3789       & 0.8760        & 0.1746       & -              & 31.3062       & 0.8015        & 0.1255       & - \\
        \textbf{LRTV\cite{7167714}}    & \cellcolor{tabthird}33.5149  & 0.9078   & 0.1566 & - & 32.7237       & 0.8561        & \cellcolor{tabthird}0.0939 & - \\
        \textbf{LRTF-DFR\cite{9084248}}& 33.3251       & \cellcolor{tabthird}0.9286 & \cellcolor{tabthird}0.1062 & -  & \cellcolor{tabthird}33.5054  & \cellcolor{tabthird}0.9059 & \cellcolor{tabfirst}0.0668 & - \\
        \textbf{SSLR-SSTV\cite{rs13040827}} & \cellcolor{tabsecond}35.9781 & \cellcolor{tabsecond}0.9674 & \cellcolor{tabfirst}0.0860 & - & \cellcolor{tabsecond}34.6840 & \cellcolor{tabfirst}0.9494 & \cellcolor{tabsecond}0.0806 & - \\
        \textbf{Ours (HSI-Diffusion)} & \cellcolor{tabfirst}38.9210 & \cellcolor{tabfirst}0.9789 & \cellcolor{tabsecond}0.0923 & 0.1531 & \cellcolor{tabfirst}36.2172 & \cellcolor{tabsecond}0.9473 & 0.0928 & 0.2213 \\
        \bottomrule
    \end{tabular}
\end{table*}

\subsection{Creating Synthetic Data}
\label{sec:synthetic}

Given the absence of real striped hyperspectral satellite data, we devise a method to synthesize such data for model training. This approach is essential for our denoising models, which rely on accurately replicating and subsequently removing common artifacts like stripes from hyperspectral images. Our methodology begins with ground-truth hyperspectral cubes which are free from any artifacts, providing a clean baseline from which noise can be methodically introduced.

To emulate realistic stripe noise, we introduce Gaussian noise stripes to the clean hyperspectral data. The stripe noise is generated as follows,
\begin{equation}
\mathbf{S}_{i}(x,y,z) = \mathbf{G}(0, \sigma^2) \cdot \text{intensity\_factor}(i), 
\end{equation}

where $\mathbf{G}(0, \sigma^2)$ represents a Gaussian noise matrix with mean 0 and variance $\sigma^2$, and $\text{intensity\_factor}(i)$ denotes the stripe intensity for the $i$-th band. The stripe intensity is calculated as a percentage of the dynamic range of each spectral band, which varies from $0.1\%$ to $5\%$. Specifically, $\text{intensity\_factor}(i)$ is determined by,
\begin{equation}
\begin{split}
    \text{intensity\_factor}(i) &= \text{max}\left(0.001, \text{stripe\_intensity}\right. \\ & - 0.05 + \mathcal{U}(0, 0.1)\left.\right), 
\end{split}
\end{equation}

where $\text{stripe\_intensity}$ is sampled from a uniform distribution $\mathcal{U}(0.01, 0.3)$ initially, and $\mathcal{U}(a, b)$ represents a uniform random variable between $a$ and $b$. Based on the stripe frequency and with uniform probability, columns are randomly selected for striping. These stripes are fragmented across the column, with the number of fragments following a uniform distribution. In addition, the stripes can vary in size and may not extend throughout the column, and the size is also uniformly distributed. Each band will have stripes generated with this method.

Following the introduction of synthetic stripe noise, we sample $32 \times 32 \times 32$ cubes from the larger hyperspectral images. These cubes are randomly sampled across spatial and spectral dimensions to ensure variability in training data, thus preparing the model to handle a wide range of real-world scenarios.

To generate additional data, we also use augmentation techniques such as CutMix~\cite{Yun_2019_ICCV} and Mixup~\cite{zhang2018mixup}, which we apply specifically in the frequency bands of the hyperspectral cubes. For each sampled hyperspectral cube, we apply the frequency-specific augmentation as follows:
\begin{equation}
\mathbf{X}_{\text{aug}}^{(f)} = \lambda \mathbf{X}_1^{(f)} + (1 - \lambda) \mathbf{X}_2^{(f)},
\end{equation}
where $f$ indicates the frequency band, $\mathbf{X}_1^{(f)}$ and $\mathbf{X}_2^{(f)}$ are two randomly chosen hyperspectral cubes, and $\lambda$ is sampled from a Beta distribution, $\text{Beta}(\alpha, \alpha)$ for CutMix and a uniform distribution $\mathcal{U}(0,1)$ for Mixup. Here, $\alpha$ is a hyperparameter to adjust the amount of mixing.

\section{EXPERIMENTS}

We also perform experiments to assess the performance of our model. We perform experiments on the ICVL-HSI~\cite{10.1007/978-3-319-46478-7_2} dataset where we evaluate our method on performing denoising. Many modern approaches evaluate on the ICVL dataset~\cite{10.1007/978-3-319-46478-7_2} by adding a synthetic Gaussian noise and we evaluate our model on this benchmark too to compare our approach with other models. While we present these results for completeness, performing well on the ICVL dataset~\cite{10.1007/978-3-319-46478-7_2} is not a goal of this work, as our method aims to solely perform well on satellite hyperspectral images. We also evaluate our method on our synthetic dataset created with the approach outlined in~\Cref{sec:synthetic} on remote sensing hyperspectral images. Furthermore, we also evaluate our method on the EnMAP datacube in the region that our FINCH mission would image that is thus comparable to our satellite's destriping needs.

All of our final code was optimized for a 1 x A100-80 GB GPU. For our Diffusion model, we trained our model for $30K$ steps per stage with a batch size of $32$ on the $32\times 32\times 32$ resolution. We use the Adam optimizer~\cite{kingma2017adam} with $\beta_1=0.9$, $\beta_2=0.999$, and $\epsilon=10^{-8}$ with an initial learning rate of $10^{-4}$. We use cosine decay for the learning rate. We use the same optimization setup throughout all of our stage, however, we scale the learning rate according to our dataset size for each of the stages.

\subsection{Quantitative Results}

\begin{figure*}[t]
    \centering
    \begin{tabular}{cccccc}
    Band 100 & Band 124 & Band 148 & Band 172 & Band 196 & Band 220 \\
    \midrule
    \includegraphics[width=0.145\textwidth]{images/enmap/gray/in_100.png} &
    \includegraphics[width=0.145\textwidth]{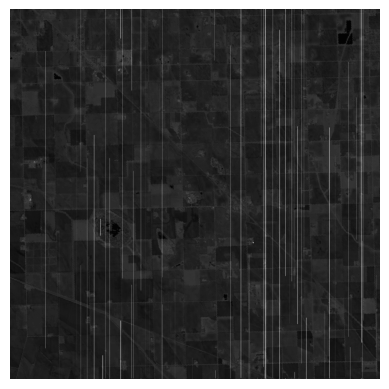} &
    \includegraphics[width=0.145\textwidth]{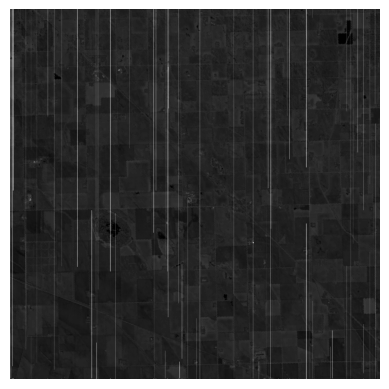} &
    \includegraphics[width=0.145\textwidth]{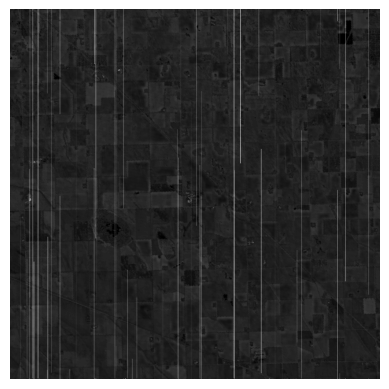} &
    \includegraphics[width=0.145\textwidth]{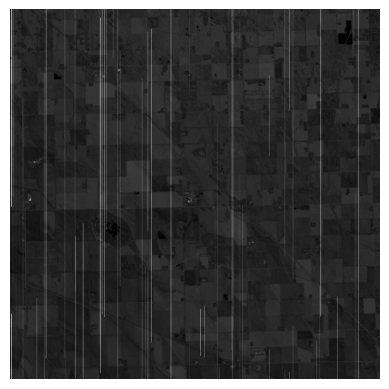} &
    \includegraphics[width=0.145\textwidth]{images/enmap/gray/in_220.png}\\
    \includegraphics[width=0.145\textwidth]{images/enmap/gray/generated_100.png} &
    \includegraphics[width=0.145\textwidth]{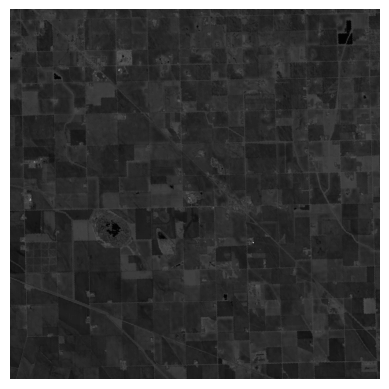} &
    \includegraphics[width=0.145\textwidth]{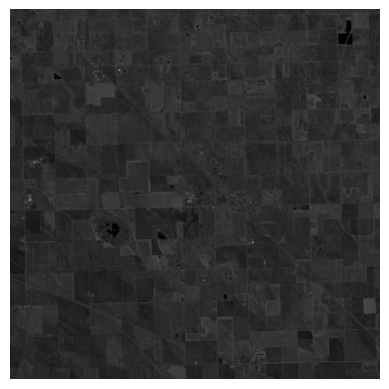} &
    \includegraphics[width=0.145\textwidth]{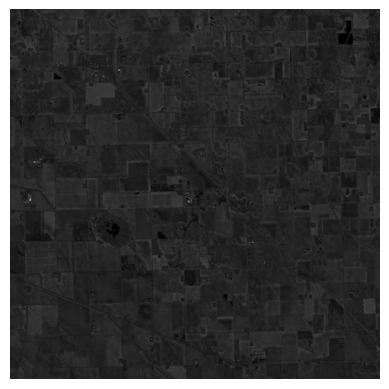} &
    \includegraphics[width=0.145\textwidth]{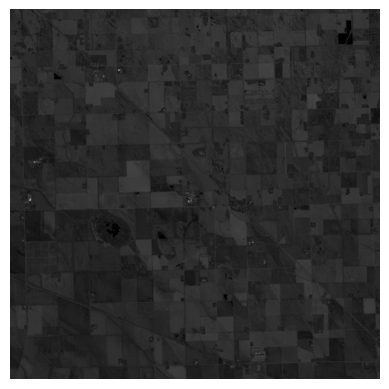} &
    \includegraphics[width=0.145\textwidth]{images/enmap/gray/generated_220.png}
    \end{tabular}
    \caption{\emph{Qualitative Results.} We present striped (Row 1) and the corresponding destriped bands produced from our methods (Row 2) from the EnMAP~\cite{STORCH2023113632, rs70708830} image of FINCH's imaging site, with stripes added as described in~\Cref{sec:synthetic}. From left to right, the images correspond to bands 100, 124, 148, 172, 196, and 220. Best viewed with zoom.}
    \label{fig:enmap-black-and-white}
\end{figure*}

\begin{figure*}[t]
    \centering
    \begin{tabular}{cccccc}
    Band 100 & Band 124 & Band 148 & Band 172 & Band 196 & Band 220 \\
    \midrule
    \includegraphics[width=0.145\textwidth]{images/enmap/jet/in_100.png} &
    \includegraphics[width=0.145\textwidth]{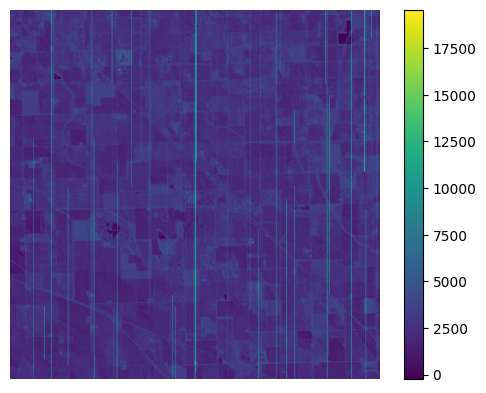} &
    \includegraphics[width=0.145\textwidth]{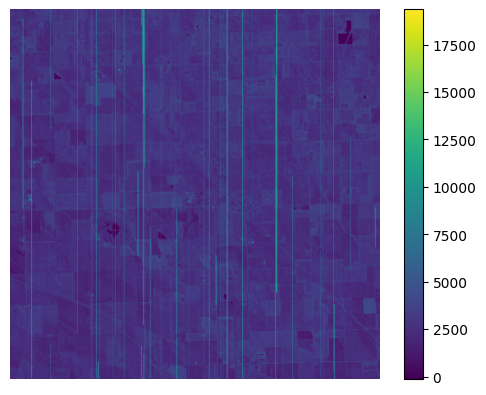} &
    \includegraphics[width=0.145\textwidth]{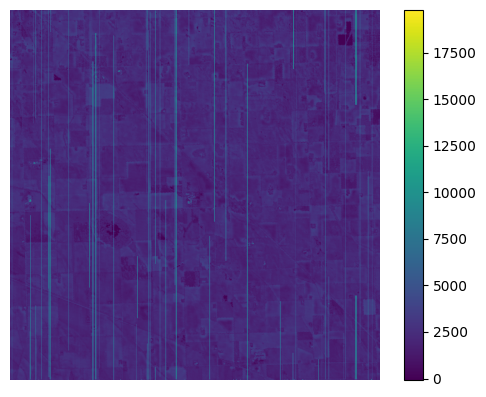} &
    \includegraphics[width=0.145\textwidth]{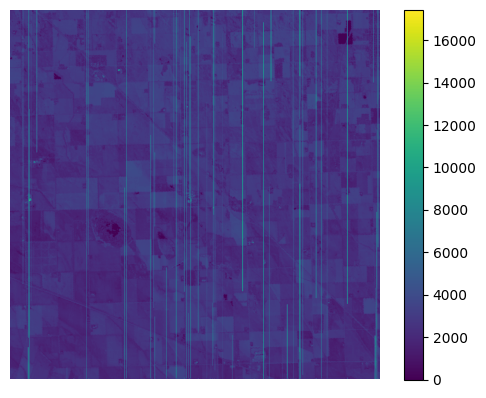} &
    \includegraphics[width=0.145\textwidth]{images/enmap/jet/in_220.png}\\
    \includegraphics[width=0.145\textwidth]{images/enmap/jet/generated_100.png} &
    \includegraphics[width=0.145\textwidth]{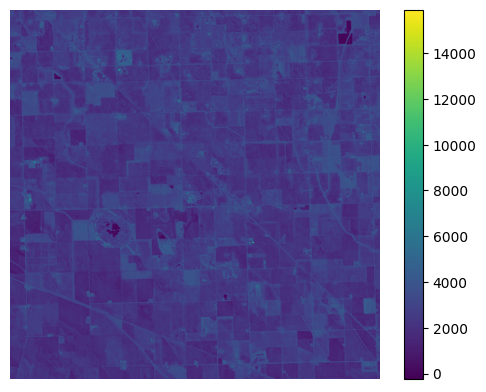} &
    \includegraphics[width=0.145\textwidth]{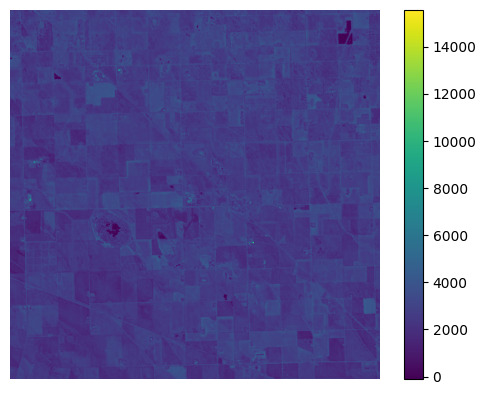} &
    \includegraphics[width=0.145\textwidth]{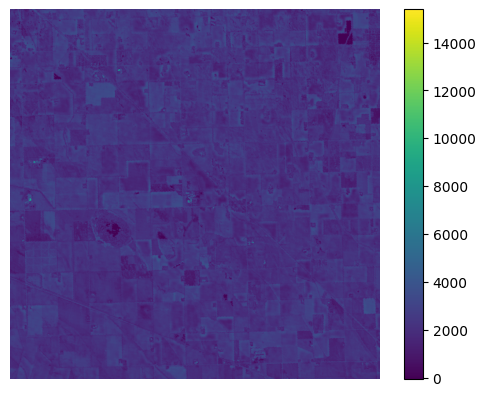} &
    \includegraphics[width=0.145\textwidth]{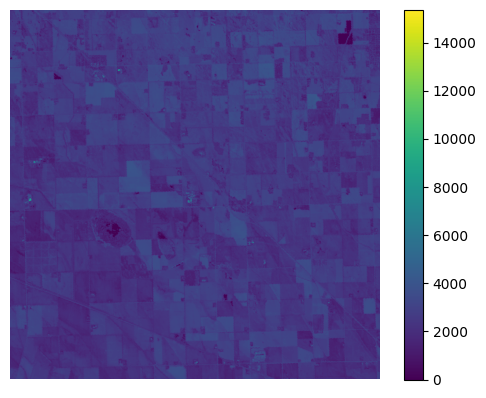} &
    \includegraphics[width=0.145\textwidth]{images/enmap/jet/generated_220.png}
    \end{tabular}
    \caption{\emph{Qualitative Results.} We present striped (Row 1) and the corresponding destriped (Row 2) bands from the EnMAP~\cite{STORCH2023113632, rs70708830} image of FINCH's imaging site, with stripes added as described in~\Cref{sec:synthetic}. From top left to bottom right, the images correspond to bands 100, 124, 148, 172, 196, and 220. We present them in the Viridis~\cite{viridis} colour scheme in this figure, with intensity bars to aid the reader in understanding the image. Best viewed in color and in conjunction with~\Cref{fig:enmap-black-and-white}.}
    \label{fig:enmap-viridis}
\end{figure*}

We demonstrate our results on the ICVL-HSI dataset~\cite{10.1007/978-3-319-46478-7_2} in~\Cref{tab:real_results} and our results on the synthetically crafted satellite hyperspectral images in~\Cref{tab:syn_results}. We evaluate our method and other approaches based on the PSNR, SSIM, spectral angle mapper (SAM)~\cite{yuhas1992discrimination}, and LPIPS~\cite{Zhang_2018_CVPR} metrics which are commonly used to evaluate denoising. While our method is unable to perform well on the ICVL-HSI dataset~\cite{10.1007/978-3-319-46478-7_2}  we report these for completeness and not as the focus of our results. We demonstrate that our approach beats state-of-the-art approaches on the synthetic dataset.

\subsection{Qualitative Results}

We present qualitative results on the EnMAP datacube~\cite{STORCH2023113632, rs70708830} from the region that would be imaged by our FINCH mission in~\Cref{fig:enmap-black-and-white} and~\Cref{fig:enmap-viridis}. These images appear consistent with the original images and maintain scene data such as field boundaries, which are important for the FINCH mission. Of critical importance to the FINCH mission is that the destriped images visually appear to maintain similar intensity values to their striped counterparts (besides the striping artifacts). This suggests that our model preserves spectral information which is critical for achieving FINCH's mission of crop residue mapping.
 
Of note for the FINCH mission is the PSNR that our model achieved on the task of destriping remote sensing data. The PSNRs of 36 and above are well above the FINCH mission's requirement of a SNR of 30 for performing scientific retrieval~\cite{miles2022finch}.

\section{DISCUSSION AND CONCLUSION}

We introduce a 3D diffusion model specifically tailored for hyperspectral remote sensing data from the FINCH spacecraft. We train the model on widely available hyperspectral data and fine-tune the training with remote sensing data synthetically striped to prepare the model for the task of destriping FINCH data. We observe favorable results in the task of destriping hyperspectral remote sensing data.

\paragraph{Limitations.}
The model has not been trained or tested on actual FINCH data, so we cannot be certain of how it will perform on said data. The model will likely need fine-tuning once FINCH data becomes available for several reasons: we expect FINCH data to have a much lower spatial resolution and larger swath width than existing datasets, and we expect that FINCH's spectral range and spectral resolution may be different from many of the publicly available datasets. Regarding the spatial resolution and swath width, we can expect that FINCH data may present different features or the same features at different scales than existing hyperspectral datasets. This may require a degree of retraining in our model. Similarly, regarding the spectral range and resolution, since many of the publicly available remote sensing datasets may cover spectral ranges that only partly overlap with FINCH's spectral range, we may see spectral features in FINCH data that were not seen during training or said features may appear at a resolution that was not seen during training. This last limitation is particularly relevant for the FINCH mission, where accurate spectral information is critical for achieving the mission's scientific objective, as mentioned earlier.

\paragraph{Future Applications.} Within the context of the FINCH mission, there are several opportunities for future work with this model. The small form factor and focal length of the FINCH spacecraft and payload motivate the requirement for a super-resolution technique to improve the spatial resolution of FINCH data for tasks such as georeferencing and imaging site identification. Additionally, different types of noise, such as dark current noise, may require more sophisticated machine learning-based denoising approaches, depending on the precision of thermal sensors employed within the final space-ready version of FINCH. Beyond our FINCH mission, our approach would be very useful for denoising hyperspectral images in general. Our hyperspectral diffusion model may lend itself well to being applied to both of these tasks in the future. Finally, future work could analyze and fine-tune this model's performance on actual FINCH data, after FINCH has been launched and has started operating.

\section*{Acknowledgements}

We gratefully acknowledge the University of Toronto Student Union and the student body for their generous funding of UTAT-Space Systems through the UTAT-Innovation Fund\footnote{\url{https://www.utsu.ca/u-of-t-service-groups/}}. Additionally, we also thank the University of Toronto Engineering Society\footnote{\url{https://skule.ca/}} for their substantial support through grants and funding.

\newpage
\bibliography{references}
\bibliographystyle{unsrt}
\end{multicols*}

\end{document}